\documentclass{aa}

\usepackage{graphicx}

\begin{document}

\title{Resistive MHD simulations of the Parker instability in galactic disks}

\author{Grzegorz Kowal\inst{1}
  \and Micha\l{} Hanasz\inst{2}
  \and Katarzyna Otmianowska-Mazur\inst{1}}

\institute{Astronomical Observatory, Jagiellonian University, ul. Orla 171, 30-244 Krak\'{o}w, Poland
  \and Toru\'{n} Centre for Astronomy, Nicholas Copernicus University, 87-148 Piwnice/Toru\'{n}, Poland}

\authorrunning{G. Kowal et al.}
\titlerunning{Resistive MHD simulations of the Parker instability...}

\offprints{Grzegorz Kowal, \email{kowal@oa.uj.edu.pl}}

\date{Received 30 September 2002 / Accepted 25 March 2003 }

\abstract{ Parker instability leads to the formation of tangential
discontinuities in a magnetic field and subsequent magnetic reconnection due to
a numerical and/or an explicit resistivity. In this paper we investigate the
role of the uniform, localized and numerical resistivity on the diffusion of
magnetic field lines during the growth phase of Parker instability modes. We
propose a new method to quantify the diffusion of magnetic field lines which is
attributed to the presence of resistivity in ideal and non-ideal MHD codes. The
method relies (1) on integration of magnetic lines in between periodic
boundaries, (2) on measurements of the dispersion of magnetic field lines with
the left and the right periodic boundaries and (3) on a statistical analysis of
shifts of a large set of magnetic lines. The proposed method makes it possible
to detect topological evolution of magnetic field. We perform a series of
resistive MHD simulations of the Parker instability in uniformly rotating
galactic disks. We follow the topological evolution of the magnetic field
evolving due to the Parker instability and relate it to the ratio of total to
uniform magnetic field in galactic disks. We find that after the onset of the
Parker instability, the magnetic field becomes first tangled and later on it
evolves toward a uniform state due to the presence of resistivity. A similar
effect of a varying contribution of a turbulent magnetic field is observed in
arms and inter-arm regions of galaxies.

\keywords{ISM: magnetic diffusion -- Galaxies: structure -- Magnetohydrodynamics
(MHD)}}

\maketitle

\section{Introduction}
\label{sec:intro}

The issue of stability of magnetized astrophysical disks is of primary
importance for their dynamical evolution. Parker (\cite{parker1},
\cite{parker2}) discovered that galactic disks containing a large scale
horizontal magnetic field and vertical disk gravity are unstable against the
buoyancy of magnetic field and cosmic rays. When the contribution of the
weightless magnetic field and cosmic rays to the total pressure is significant,
small perturbations of the system grow in amplitude, form wavelike perturbations
of magnetic field lines and promote sliding of the heavy gas to magnetic
valleys. The typical horizontal wavelength of unstable modes is of the order of
a few hundreds parsecs up to a few kiloparsecs. This kind of instability
transports the magnetic field out of the disks. On the other hand Parker
instability is supposed to contribute to the magnetic field amplification
through the process of fast galactic dynamo introduced originally by Parker
(\cite{parker3}) and developed subsequently in a series of papers by Hanasz \&
Lesch (\cite{hanasz1}, \cite{hanasz2}, \cite{hanasz3}) and Hanasz,
Otmianowska-Mazur \& Lesch (\cite{hanasz4}, hereafter referred to as Paper I),
and by Brandenburg and Schmitt (\cite{brandenburg}) and Moss et
al. (\cite{moss}).

Another relevant MHD instability is the magneto-rotational instability, referred
to as Balbus-Hawley instability (\cite{balbus1}, \cite{balbus2}), which relies
on the angular momentum exchange, by means of perturbed magnetic field lines,
between different orbits of differentially rotating magnetized gaseous disks.
The development of this instability initiates and sustains turbulence and leads
to the amplification of the small scale magnetic field up to a certain level of
saturation. The common feature of both instabilities is the development of
non-uniformities and contact discontinuities on the initially smooth magnetic
fields.

According to Ampere's law, magnetic field discontinuities correspond to current
sheets. Currents are dissipated according to Ohm's law and magnetic energy is
converted into the thermal energy. As a consequence magnetic field is no longer
frozen in the gas flows and the magnetic field topology can evolve due to a
nonidealness of the medium. Resistive processes may influence the development of
MHD instabilities in astrophysical conditions. The issue has been undertaken in
both cases of Balbus-Hawley (Fleming et al. \cite{fleming}) and Parker
instabilities (Paper I).

A significant part of studies of MHD instabilities in astrophysics have been
done with the aid of numerical simulations. The intrinsic property of numerical
codes solving MHD equations is the presence of a code-dependent numerical
diffusivity. Therefore, even if the ideal set of equations is actually being
solved, the numerical diffusion of the magnetic field changes the magnetic field
topology. In view of the above facts, the main goal of the present project is to
analyze the influence of the different kinds of resistivity on the magnetic
field topology in numerical simulations of the Parker instability.

We apply three types of a resistivity: the uniform one, the localized
resistivity operating only above some critical value of the current density, and
the numerical one. Our model study is based on the general setup similar to that
applied in Paper I, i.e. we consider a simple equilibrium state in a uniform
gravity, assuming that magnetic pressure is initially proportional to the gas
pressure. The initial magnetic field is vertically stratified, and is parallel
to the azimuthal direction. We superpose a perturbation of velocity and simulate
the Parker instability in a local Cartesian box with coordinates $(x,y,z)$
parallel locally to the directions of $r$, $\phi$ and $z$ coordinates. We take
into account the disk rotation introducing the Coriolis force. We apply the
resistivity model following Paper I (and references therein).

In order to perform a quantitative estimation of the effect of resistivity in a
Parker unstable system we introduce a new method of measuring the topological
variations of magnetic field lines. The method described in Section
\ref{sec:results} allows us to examine the effects of different prescriptions of
the resistivity and to compare them to the effect of the numerical resistivity.
The applied concept is similar to the description of magnetic diffusion with the
aid of Liapunov exponents (Barge et al., \cite{barge}, Barghouty and Jokipii
\cite{barghouty}, Zimbardo \cite{zimbardo}). Our method is different from the
one applied by Fleming et al. (\cite{fleming}), where the resistivity is
measured through the determination of the saturation of the growth of magnetic
energy as the result of the Balbus-Hawley instability. In the case of Parker
instability such a method does not provide any information about the resistivity
since without the action of differential rotation we do not observe any growth
of the magnetic energy.

We are primarily interested in the role of the resistivity in uniformization of
the magnetic field structure distorted by the onset of the Parker
instability. Referring to the picture of the galaxy NGC 6946 presented by Beck
and Hoernes (\cite{beck}) we note that highly unpolarized synchrotron emission
comes from the regions of optical arms and a polarized emission comes from
inter-arm regions. The reason for the depolarization in arms is obviously a
strong agitation of the interstellar medium (ISM) by stellar explosions and
winds. These phenomena form a very intense source of perturbations which are
supposed to contribute to the excitation of the Parker and other instabilities.
Since density waves are interpreted as waves traveling across the galactic disk
area (Binney and Tremaine \cite{binney}) optical arms were placed in the past in
the present location of inter-arm regions. One can ask therefore which mechanism
is responsible for the converting of the turbulent medium in the arms in the
much more uniform medium in the inter-arm regions. Within the framework of our
model, we can suggest that the magnetic reconnection can play an important role
in the uniformization of galactic magnetic fields.

\section{Method}
\label{sec:method}

\subsection{Equations}
\label{sec:method_equations}

We investigate the Parker instability in galactic disk in the presence of a
rigid rotation and a fast magnetic reconnection assuming isothermal evolution of
the system following Paper I.

We solve a set of MHD equations including the Coriolis force in the equation of
motion and the resistive term in the induction equation:
\begin{equation}
\frac{\partial \rho}{\partial t} + \nabla \cdot \left( \rho \vec{v} \right) = 0
\, ,
\label{eqn:mhd_cont}
\end{equation}
\begin{equation}
\frac{\partial \vec{v}}{\partial t} + \left( \vec{v} \cdot \nabla \right)
\vec{v} = - \frac{1}{\rho} \nabla \left( p + \frac{B^2}{8 \pi} \right) + \frac{
\left( \vec{B} \cdot \nabla \right) \vec{B}}{4 \pi \rho} - 2 \vec{\Omega} \times
\vec{v} + \vec{g} \, ,
\label{eqn:mhd_motion}
\end{equation}
\begin{equation}
\frac{\partial \vec{B}}{\partial t} = \nabla \times \left( \vec{v} \times
\vec{B} - \eta \, \vec{j} \right) \, ,
\label{eqn:mhd_induc}
\end{equation}
where $\vec{\Omega}$ is the angular rotation velocity, $\vec{j} = \nabla \times
\vec{B}$ is the current density, $\vec{g}=g \, \hat{e_{\rm z}}$, $g={\rm const}$
is the gravitational acceleration and the other symbols have their usual
meaning. Our assumption of uniform gravity is not realistic for actual galaxies,
nevertheless it is satisfactory for a theoretical investigations of the Parker
instability.

We adopt an isothermal equation of state
\begin{equation}
p = c_{\rm s}^2 \rho \, ,
\end{equation}
where $c_{\rm s}{\rm = const}$ is the isothermal sound speed.

The following Ansatz for an anomalous resistivity (Ugai \cite{ugai}, Konz et al.
\cite{konz}, Tanuma et al. \cite{tanuma}, Paper I) is applied:
\begin{equation}
\eta (j) = \eta_1 + \eta_2 \left( j^2 - j_{\rm crit}^2 \right)^{1/2} \Theta
\left( j^2 - j_{\rm crit}^2 \right) \, ,
\label{eqn:anom_resis}
\end{equation}
where $\eta_1$ and $\eta_2$ are constant coefficients, $j_{\rm crit}$ denotes
the critical current density above which the resistivity switches on and
$\Theta$ is the Heaviside step function. Such a prescription of the resistivity
allows for a useful parametrization of the magnetic reconnection, as discussed
in Paper I.

Following Paper I we introduce the Coriolis force corresponding to the rotation
angular velocity $\Omega$ which takes a typical galactic value defined in
section 2.3. The Coriolis force is responsible for horizontal deflections of
magnetic field lines, which are smaller in magnitude than the vertical
deflections. The role of the Coriolis force is to twist rising magnetic loops
against galactic rotation and to produce, in cooperation with vertical gravity
and resistivity, the helical magnetic field structures apparent in simulations
of Paper I.

In order to integrate MHD equations (\ref{eqn:mhd_cont}) - (\ref{eqn:mhd_induc})
we apply the ZEUS-3D code (Stone \& Norman \cite{stone1}, \cite{stone2}) with
our own modifications necessary to introduce the Coriolis force and to
incorporate the localized resistivity (see Paper I for more details).

\subsection{The method of measuring of the effect of the resistivity}
\label{sec:method_measure}

The presence of the resistivity in the induction equation releases the magnetic
freezing condition and leads to a topological evolution of magnetic field lines.

\begin{figure}[t]
  \centering
  \resizebox{0.95\columnwidth}{!}{\includegraphics{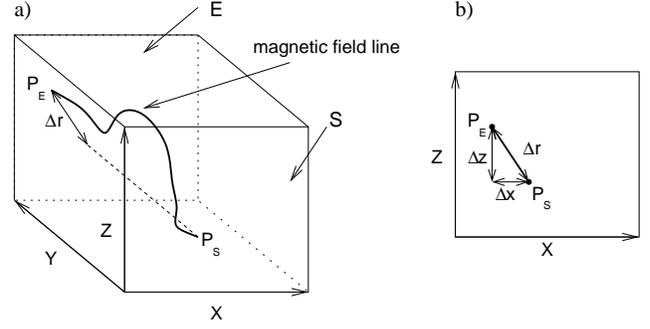}}
  \caption{Figure describing method of measuring the dispersion of magnetic
    field lines due to resistivity. $S$ and $E$ are boundary plane at beginning
    and ending of $y$ coordinate. $\mathrm{P_S}$ and $\mathrm{P_E}$ are points on
    the XZ plane where magnetic line starts and ends. $\Delta \mathrm{x_i}$,
    $\Delta \mathrm{z_i}$, $\Delta \mathrm{r_i}$ is a deviation between the
    starting and the ending points of the $i$-th line. The right panel presents
    the projection of the deviation on the XZ plane.}
  \label{fig:method}
\end{figure}

Measurements of deviations of a single magnetic field line from the perfect
freezing in gas motion is possible through integration of magnetic vector field
in a 3D Cartesian volume bounded by periodic boundaries. We propose a method
that is illustrated in Fig. \ref{fig:method}. We assume that the initial
magnetic field is parallel to the Y-axis in the whole volume of the
computational box. We apply periodic boundary conditions on the XZ as well as
YZ boundaries. We integrate magnetic field lines between the starting XZ
boundary denoted by the symbol "S" and the ending XZ boundary denoted by "E".
We consider a family of $N$ magnetic field lines. Each member of the family is
denoted by the index $i$. Our choice of initial magnetic field parallel to the
Y direction ensures that magnetic field lines are closed by periodic
boundaries, i.e. the intersection points $(X_i,Z_i)_S$ of the $i$-th line in the
S plane are identical as the intersection points $(X_i,Z_i)_E$ in the E plane.

In the case of ideal MHD the magnetic freezing condition means that a given
element of the fluid will remain on the same magnetic flux tube for an
arbitrarily long period of time. This means that if $(X_i,Z_i)_S = (X_i,Z_i)_E$
at $t=0$ then $(X_i,Z_i)_S = (X_i,Z_i)_E$ at any later time (instant), no matter
what the evolution of the box interior is. If, on the other hand, the system is
non-ideal i.e. resistivity is present, then magnetic field lines become open
$(X_i,Z_i)_S \neq (X_i,Z_i)_E$ at $t > 0$. We define deviations of the
intersection points of magnetic lines with the S and E boundaries, as $(\Delta
X_i,\Delta Z_i) = (X_i,Z_i)_E - (X_i,Z_i)_S$ (see Fig.~\ref{fig:method}).

As soon as $(\Delta X_i,\Delta Z_i)$ are known as functions of time one can
perform a statistical analysis through the computation of the mean and the
standard dispersion of deviations of the E and S intersection points.
\begin{eqnarray}
\overline{\Delta x} &=& \frac{1}{N} \sum_{i=1}^N \Delta x_{i},\\
\overline{\Delta z} &=& \frac{1}{N} \sum_{i=1}^N \Delta z_{i},
\end{eqnarray}
\begin{eqnarray}
 \sigma_{x} &=& \left( \frac{1}{N-1} \sum_{i=1}^N{(\Delta x_{i} 
    - \overline{\Delta x})^2}\right)^{1/2} \, ,\\
 \sigma_{z} &=& \left( \frac{1}{N-1} \sum_{i=1}^N{(\Delta z_{i} 
    - \overline{\Delta z})^2}\right)^{1/2} \, .   
\end{eqnarray}

In the rest of the paper, for the sake of brevity, this kind of dispersion will
be referred to as the dispersion of magnetic field lines. In actual
implementation of the method of determination of the magnetic diffusion the
integration of magnetic field lines is performed with the aid of the Runge-Kutta
method of $4^\mathrm{th}$ order. This procedure is applied for a large sample of
staring points ${\rm P_S}$ on the first XZ boundary for different timesteps of
evolution of the Parker instability.

The time evolution of the statistical moments depends primarily on values of the
resistivity coefficients $\eta_1$ and $\eta_2$, as well as on the grid
resolution adopted for the MHD simulations. The presented method makes it
possible to determine the intrinsic numerical resistivity of the MHD code by
means of comparison of the dispersion measure of magnetic field lines for low
resolution runs with the vanishing resistivity and high resolution runs with
different values of the resistivity coefficient $\eta_1$. Results of this
procedure will be presented in Section \ref{sec:results}. Finally we would like
to stress out that the periodic boundary conditions are essential for our method
because we measure the departures of magnetic field lines from the closure (by
periodic boundaries). This measure indicates topological changes of field lines
and therefore the presence of non-ideal effects.

\subsection{Initial conditions}
\label{sec:method_conditions}

Our initial equilibrium state is an exponentially stratified isothermal disk
(Parker \cite{parker1}, Paper I). The initial equilibrium is characterized by an
uniform vertical gravity with the vertical gravitational acceleration $g = -2
\cdot 10^{-9} \, \mathrm{cm \, s^{-1}}$, the midplane gas density $n_0= 1 \,
{\rm cm}^{-1}$ and the sound speed $c_{\rm s} = 7\, \mathrm{km \, s^{-1} }$,
constant across the disk. The fixed ratio of the magnetic pressure to the gas
pressure is given by $\alpha = p_{\rm mag} / p_{\rm gas}$, so that the total
pressure is given by $p_{\rm tot} = (1+\alpha) p_{\rm gas}$. We assume that
$\alpha = 1.0$, corresponding to the magnetic field strength of $4.5 \, {\rm \mu
G}$ at the galactic midplane. The initial magnetic field is parallel to the
azimuthal Y-axis. The dependence of equilibrium quantities on $z$ is given by

\begin{equation}
\frac{n_0(z)}{n_0(0)} = \frac{p_0(z)}{p_0(0)} =
\frac{B_0^2(z)}{B_0^2(0)} = \exp \left( - \frac{z}{H} \right) \, ,
\label{eqn:init_state} 
\end{equation} 
where $H = \left( 1 + \alpha \right) \, c_{\rm s}^2 / |g| \approx 159 \, {\rm
pc}$ the vertical scale-height. We apply the angular velocity of the galactic
rotation $\Omega = 0.025 \, {\rm Myr}^{-1}$ corresponding to that of the Solar
orbit.

In order to excite the Parker instability we apply initial perturbations of
vertical velocity component in the following form
\begin{equation}
v_{\rm z} 
= \left\{\begin{array}{lll}
   v_0 \cos \left( \frac{2 \pi n_{\rm x} x}{L_{\rm x}}\right) 
       \cos \left( \frac{2 \pi n_{\rm y} y}{L_{\rm y}} \right) 
       \sin \left( \frac{\pi z}{H} \right) & 
         {\rm for}\hspace{0.2cm} |z| \leq H\cr
  0  & {\rm elsewhere}  \cr
\end{array} \right.
\label{eqn:perturb}
\end{equation} 
where $n_{\rm x}$, $n_{\rm y}$ are the numbers of the harmonic components of the
X, Y directions, respectively. $L_{\rm x}$, $L_{\rm y}$ are the physical
sizes of the simulation domain and the initial velocity amplitude is $v_0 = 1.0
\, {\rm km s^{-1}}$. Following Paper I the initial velocity perturbations are
nonvanishing only up to the height $H$ above the disk midplane in order to take
into account the fact that the perturbations originate from the disk activity,
i.e SN explosions, stellar winds, etc. For the sake of simplicity we perturb
only the vertical velocity component.

\subsection{Numerical setup and input parameters}
\label{sec:method_parameters}

We analyze a set of numerical simulations in a domain of physical sizes of
$L_{\rm x} = 600 \, \mathrm{pc}$, $L_{\rm y} = 1800 \, \mathrm{pc}$, $L_{\rm z}
= 1200 \, \mathrm{pc}$ in the Cartesian reference frame $x$, $y$, $z$
corresponding locally to the radial, azimuthal and vertical galactic
coordinates. Given the fact that the scaleheight of the disk is 159 pc, the
density contrast between the bottom and the top of the computational box is
about 1900.

The initial perturbations follow the formula (\ref{eqn:perturb}) with $n_{\rm x}
\times n_{\rm y} = 3 \times 3$ in all cases. The periodic boundary conditions
are applied at all boundaries perpendicular to the galactic plane (at the
XZ planes and the YZ planes). At the lower and upper boundary parallel to the
galactic plane (the XY planes) the reflecting boundary conditions are applied.
We note that the reflecting boundary conditions imposed on horizontal boundaries
are formally not compatible with the vertical stratification of $B_{\rm
y}$. However, this choice of boundary conditions allows for a stable evolution
of the system on long timescales without any significant contribution of
numerical artifacts. Moreover, the integrated magnetic field lines are located
in the middle range of the full height of the computational box, i.e. the
starting points of magnetic field lines span the range $240 {\rm pc} \leq z \leq
960 {\rm pc}$ out of the full height of the computational box.

In our study we adopt several different values of the resistivity parameters
$\eta_1$, $\eta_2$ and two different grid resolutions: 30x90x60 (low) and
60x180x120 (high) as listed in Table \ref{tab:models}. We assume ${\rm pc^2 \,
Myr^{-1}}$ as the unit of $\eta$.

\begin{table}[t]
 \caption{\small Input parameters for computed models.}
 \begin{center}
  {\small
  \begin{tabular}{|c|c|r|r|r|}
   \hline Model & resolution & $\eta_1$ & $\eta_2$ & $j_{\rm crit}$ \\ \hline \hline
   p0 &  low &   0.0 &   0.0 &  -- \\
   p1 &  low &   0.0 &   0.1 & 0.1 \\
   p2 &  low &   0.0 &   1.0 & 0.1 \\
   p3 &  low &   0.0 &  10.0 & 0.1 \\
   p4 &  low &   0.0 & 100.0 & 0.1 \\
   q0 &  low &   0.1 &   0.0 &  -- \\
   r0 &  low &   1.0 &   0.0 &  -- \\
   s0 &  low &  10.0 &   0.0 &  -- \\
   t0 &  low & 100.0 &   0.0 &  -- \\ \hline
   P0 & high &   0.0 &   0.0 &  -- \\
   P2 & high &   0.0 &   1.0 & 0.1 \\
   P3 & high &   0.0 &  10.0 & 0.1 \\
   P4 & high &   0.0 & 100.0 & 0.1 \\
   Q0 & high &   0.1 &   0.0 &  -- \\
   R0 & high &   1.0 &   0.0 &  -- \\
   S0 & high &  10.0 &   0.0 &  -- \\
   T0 & high & 100.0 &   0.0 &  -- \\ \hline
  \end{tabular}
  }
 \end{center}
 \label{tab:models}
\end{table}

\subsection{Statistical analysis of the simulation results} 
\label{sec:statistics}
For each model displayed in Table \ref{tab:models} we compute a time evolution
of the following six quantities:
\begin{enumerate}

\item[(a)] the maximum of vertical velocity $v_{\rm z \, max}$ representing the
amplitude of unstable Parker modes,

\item[(b)] the ratio of mean energy densities of total and uniform magnetic
fields in the disk ($z\leq H$), $U_{\rm Bt}/U_{\rm Bu}=<B^2>/<B>^2$. The
averaging denoted by $<...>$ is performed over a volume between $z=0$ and $z=H$,

\item[(c)] the maximum resistivity $\eta_{\rm max}$ in the computational volume,

\item[(d)] the maximum of current density $j_{\rm max}$ in the computational
volume,

\item[(e)] the dispersion $\sigma_x$ of magnetic field lines in the
X direction,
\item[(f)] the dispersion $\sigma_z$ of magnetic field lines in the
Z direction.
\end{enumerate}

\section{Results}
\label{sec:results}

All simulations presented in this paper start from the same initial state as
described in Sect. \ref{sec:method_conditions}. The basic difference between
different considered models listed in Table \ref{tab:models} follows from the
choice of different values of resistivity parameters $\eta_1$ and $\eta_2$. The
identical initial conditions lead to initially identical evolution, however the
differences become apparent as soon as resistivity starts to play a role.

\begin{figure*}
  \centering
  \resizebox{0.7\hsize}{!}{\includegraphics{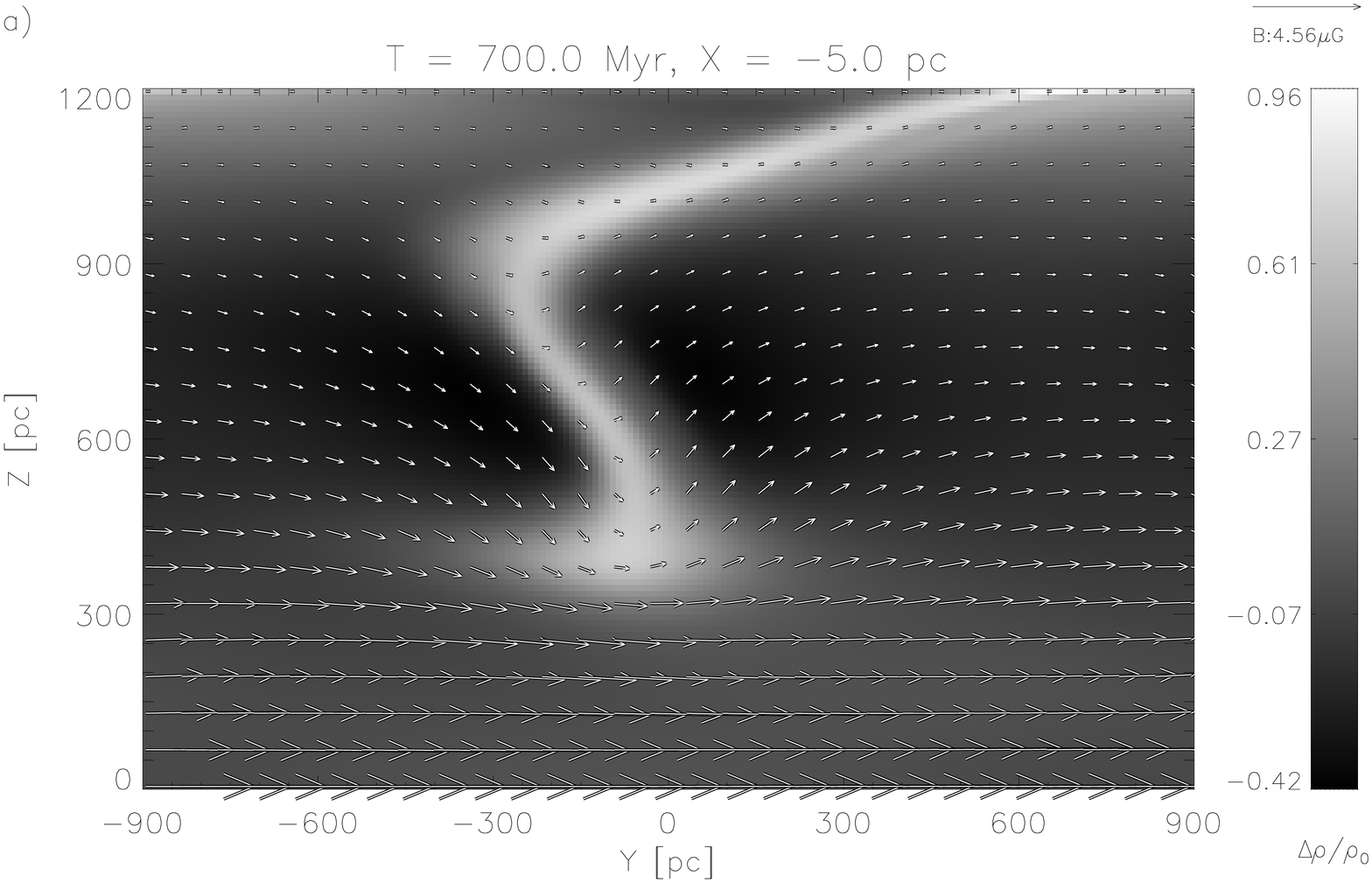}}
  \resizebox{0.7\hsize}{!}{\includegraphics{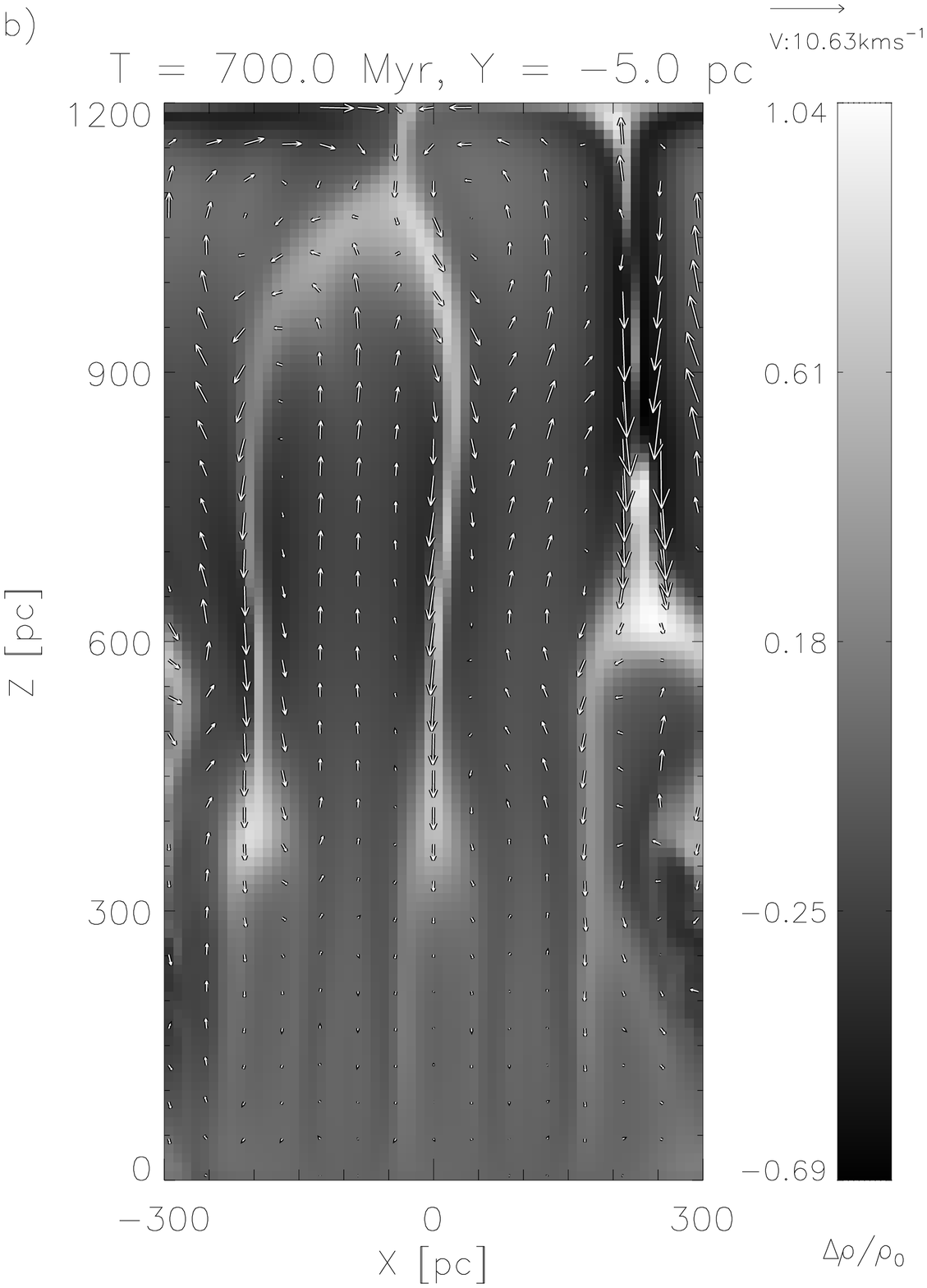}
                           \includegraphics{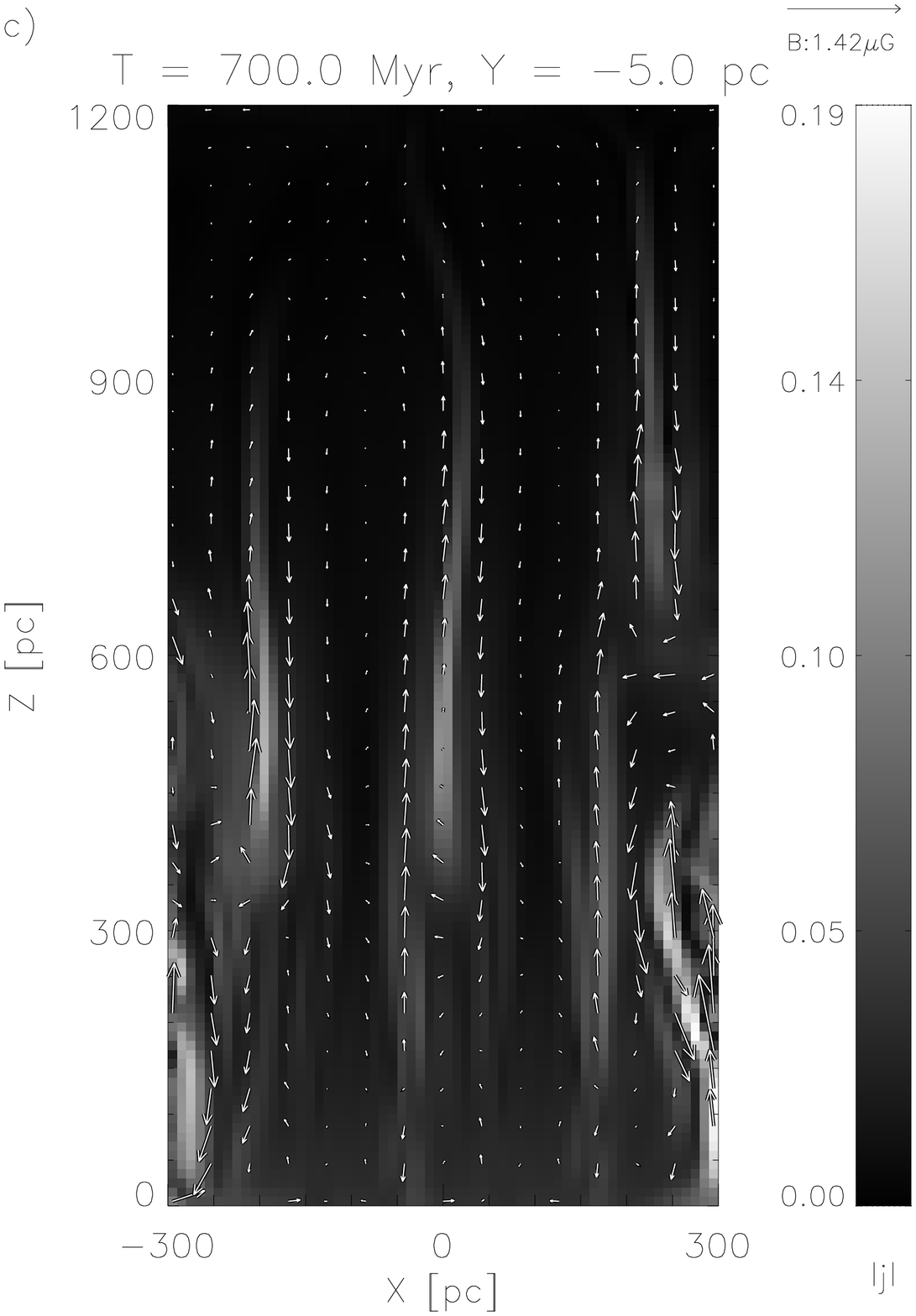}}
  \caption{Cross-sectional slices of the computational domain 
   at $t = 700$ Myr corresponding to the end of the exponential growth phase. 
   Panel (a) shows magnetic vectors and  the greyplot of density perturbation 
   $\Delta \rho/\rho_0$ in the YZ plane, panel (b) shows velocity vectors and 
   the density perturbation in the XZ plane and panel (c) shows magnetic
   vectors and current density in the same plane.}
  \label{fig:slices}
\end{figure*}

An example of cross-sectional slices at the end of exponential growth phase for
the "ideal" model p0 ($\eta_1 = \eta_2=0$) is shown in Fig. \ref{fig:slices}.
Panel (a) displays magnetic field vectors on the background of density
perturbation $\Delta \rho/\rho_0$, where $\rho_0$ is the density distribution at
$t=0$. The initially horizontal magnetic field lines are deformed in the
familiar way by an undulatory mode of the Parker instability. Gas condensation
of magnitude $\Delta \sim \rho_0$ is forming in the magnetic valley. Panel (b)
shows the density perturbation and velocity vectors in the XZ plane. It is
remarkable that the falling gas forms dense vertical sheets between rising
magnetic lobes. Panel (c) displays the XZ-components of magnetic vectors
superposed onto the current density greyplot. As it is apparent, the current
sheets form between rising magnetic lobes and coincide with the dense vertical
gas sheets. The cross-sectional slices for other models (except the cases of
large values of uniform resistivity) remain very similar to the presented one at
time T = 700 Myr.

We refer the reader to Paper I for more a detailed discussion of different
initial configurations, which include cases both with and without the explicit
resistivity. As it has been shown in Paper I, in the presence of explicit
localized resistivity reconnection starts to operate in current sheets. Before
the onset of reconnection, Parker instability deforms the initially horizontal
magnetic field into magnetic arches which are additionally twisted by the
Coriolis force. The reconnection process transforms the twisted arches into
helical magnetic flux tubes. The adjacent tubes undergo next phase of
reconnection leading to their coalescence. The final states contain a
significant radial magnetic field component, which is produced as a result of
Parker instability, the Coriolis force and magnetic reconnection.

In the absence of explicit resistivity our system starts to reconnect as well,
at later times, as a result of numerical resistivity. A comparison between
cross-sectional slices for two cases, with and without the explicit resistivity
is presented in Figs. 2(b) and 3(a) of Paper I.

In order to present the results of our statistical analysis of the perturbed
configuration we display the six statistical quantities (a)-(e), described in
Subsection \ref{sec:statistics}, in Figs.~\ref{fig:low_eta1} -
\ref{fig:high_eta2}. The plots in Figs.~\ref{fig:low_eta1} and
\ref{fig:high_eta1} show the evolution of these quantities for experiments
listed in Table~1. i.e. for different values of uniform resistivity coefficient
$\eta_1$ and $\eta_2=0$. The results for low resolution runs p0, q0, r0, s0, t0
are presented in Fig.~\ref{fig:low_eta1} and the results for the high resolution
runs P0, Q0, R0, S0, T0 are presented in Fig.\ref{fig:high_eta1}. Next two
figures present results for different values of $\eta_2$ (the localized
resistivity coefficient) and $\eta_1=0$. The results for low resolution runs p0,
p1, p2, p3, p4 are presented in Fig.~\ref{fig:low_eta2} and the results for high
resolution runs P0, P1, P2, P3, P4 are presented in Fig.~\ref{fig:high_eta2}.

\subsection{Models with different values of the coefficient $\eta_1$ }
\label{sec:results_uniform}

\begin{figure*}
  \centering
  \resizebox{0.9\hsize}{!}{\includegraphics{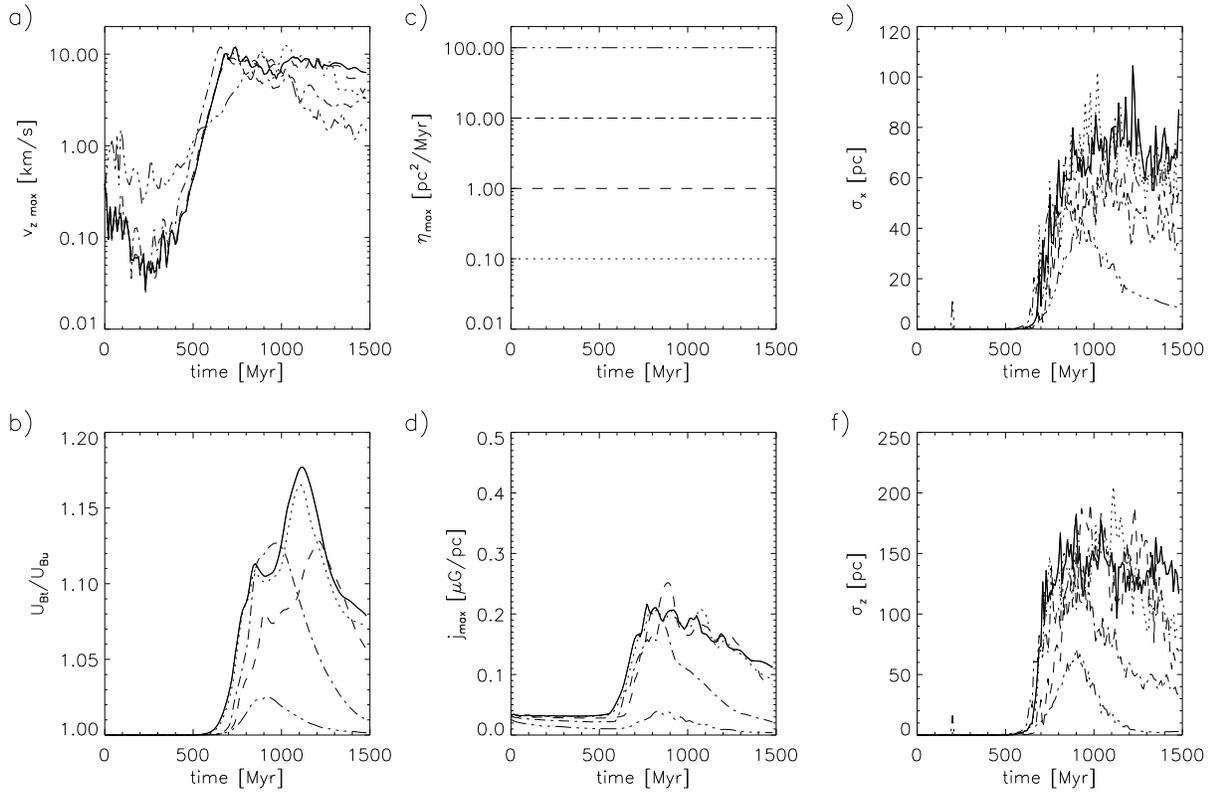}}
  \caption{Evolution of statistical quantities for low resolution. Dependence on
  $\eta_1$. (solid - p0, dotted - q0, dashed - r0, dash dot - s0, dash dot dot -
  t0)}
  \label{fig:low_eta1}
\end{figure*}

\begin{figure*}
  \centering
  \resizebox{0.9\hsize}{!}{\includegraphics{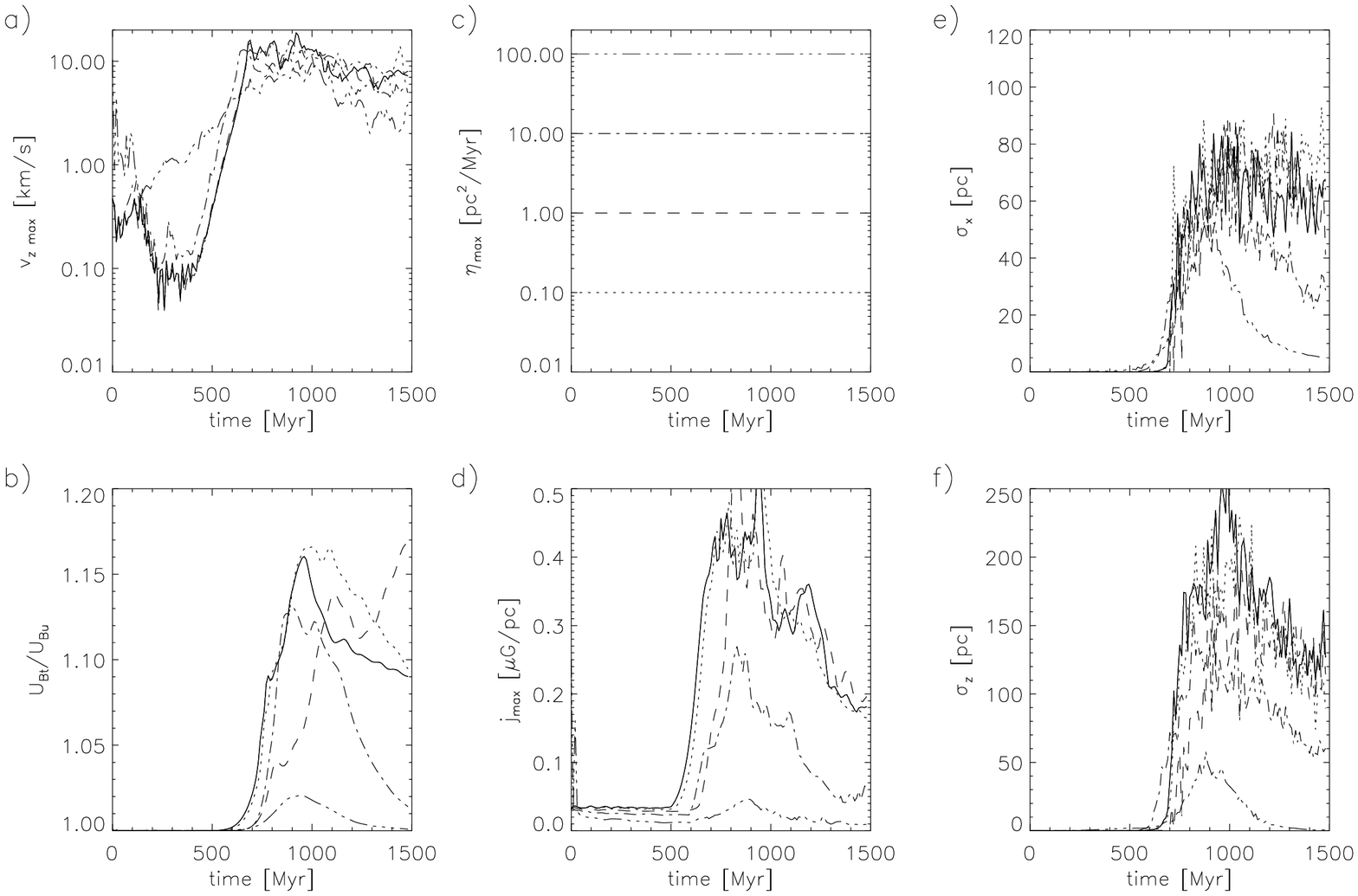}}
  \caption{Evolution of statistical quantities for high resolution. Dependence on
  $\eta_1$. (solid - P0, dotted - Q0, dashed - R0, dash dot - S0, dash dot dot -
  T0)}
  \label{fig:high_eta1}
\end{figure*}

\begin{figure*}
  \centering
  \resizebox{0.9\hsize}{!}{\includegraphics{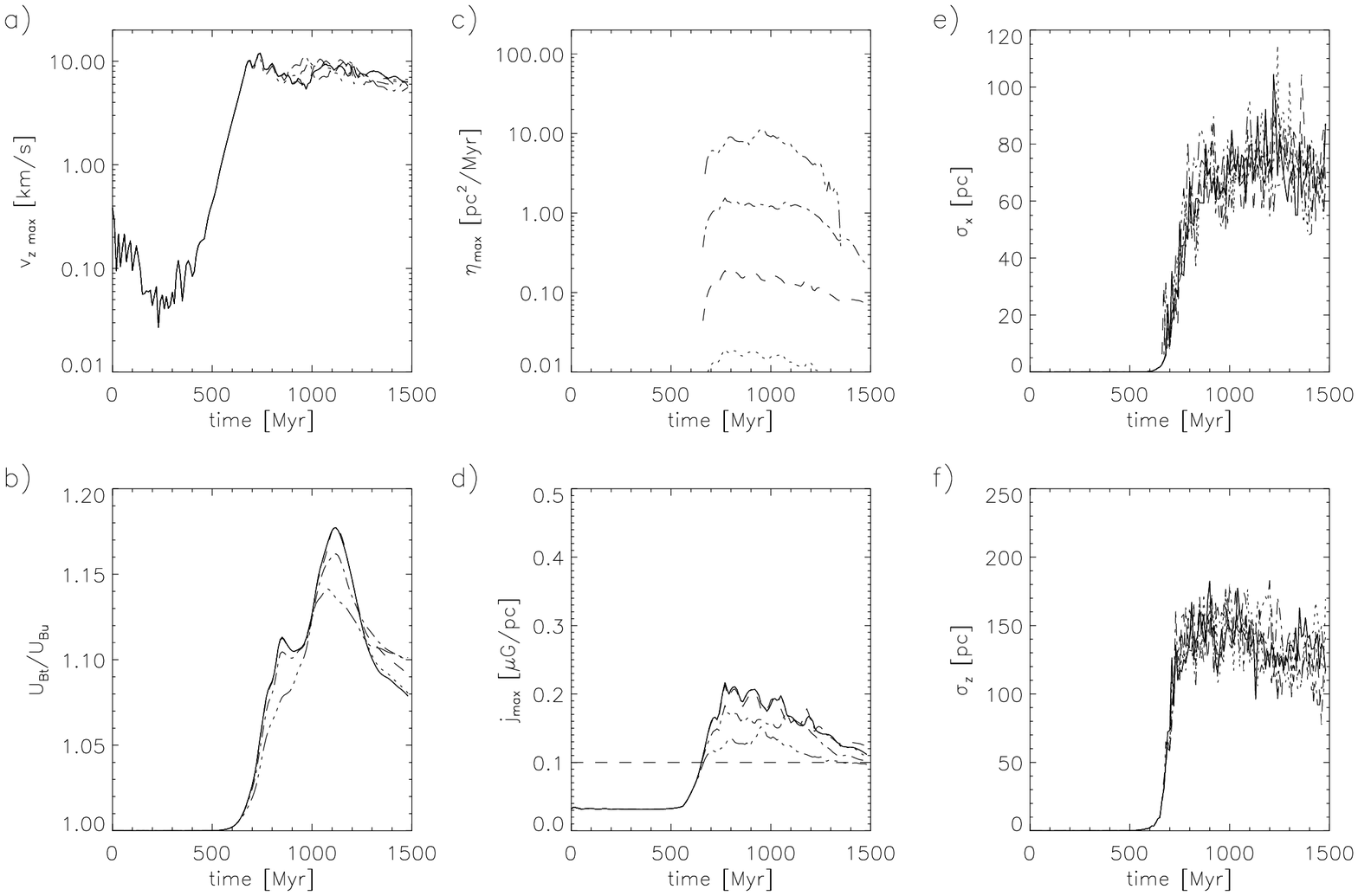}}
  \caption{Evolution of statistical quantities for low resolution. Dependence on
  $\eta_2$. (solid - p0, dotted - p1, dashed - p2, dash dot - p3, dash dot dot -
  p4)}
  \label{fig:low_eta2}
\end{figure*}

\begin{figure*}
  \centering
  \resizebox{0.9\hsize}{!}{\includegraphics{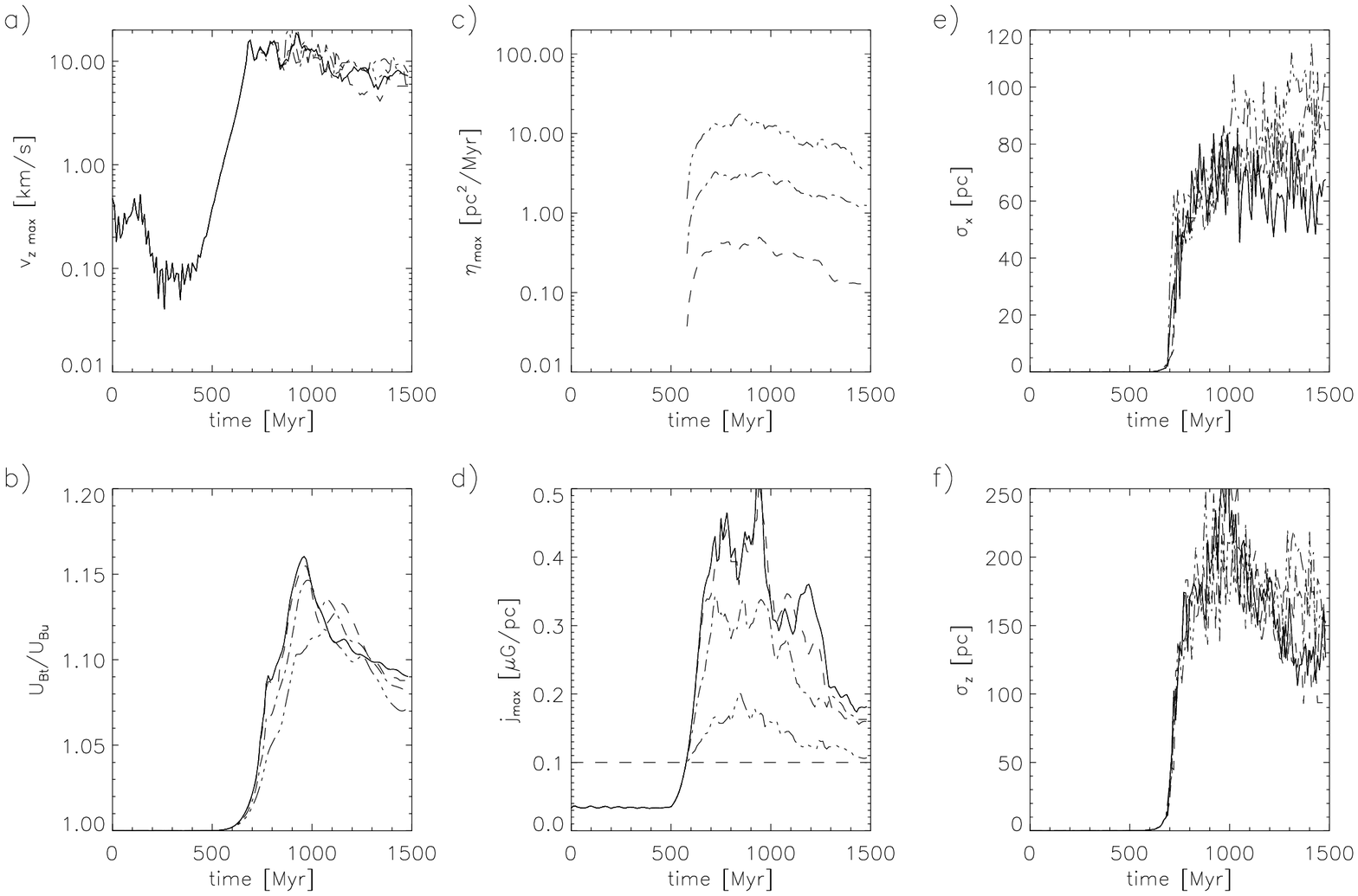}}
  \caption{Evolution of statistical quantities for high resolution. Dependence
  on $\eta_2$. (solid - P0, dotted - P1, dashed - P2, dash dot - P3, dash dot
  dot - P4)}
  \label{fig:high_eta2}
\end{figure*}

In general, all statistics are strongly dependent on the value of the
resistivity coefficient $\eta_1$ (except the maximum resistivity, which is not
representative for the present set of models). The evolution of $v_{\rm z \,
max}$ presented in Figs.~\ref{fig:low_eta1}(a) and \ref{fig:high_eta1}(a)
demonstrates a common existence of three phases: the initial phase, the
exponential growth phase and the saturation one.

During the initial phase, $v_{\rm z \, max}$ decreases because the initial
perturbation is a mixture of stable, decaying and unstable modes. The stable
modes are quickly radiated out from the perturbation region, the decaying modes
are dumped out and then the standing waves of unstable modes start to dominate
at the end of the initial phase.

The second phase of exponential growth, belonging to the linear phase of the
growth of the Parker instability, starts at about 400 Myr and lasts about 250
Myr. This period is characterized by an exponential rise of the $v_{\rm z \,
max}$ by factor of about 100 (from 0.1 to $10.0\, \rm{km s^{-1}}$). The
inclination of the curve in this phase gives a good qualification of the
evolution growth rate. For models p0, q0, r0, s0, P0, Q0, R0, S0 ($\eta_1 = 0.0
\div 10.0$), the growth rate is about $0.036$, but for the cases t0 and T0 with
$\eta_1 = 100.0$ it is only $0.009$. The maximum of $v_{\rm z}$ reaches a level
of about $10.0\, \rm {km s^{-1}}$, comparable to the Alfv{\`e}n speed, at the
beginning of the saturation phase and does not exceed it later on. For the
largest value of $\eta_1$ the evolution looks different, however we can still
distinguish three phases. Values of $v_{\rm z \, max}$ in initial phase are
larger, the growth rate is smaller, as mentioned earlier, and in the saturation
phase, the maximum velocity value decreases faster (but only for the low
resolution). In this case the presence of high resistivity apparently slows down
the linear phase of growth of the Parker instability.

Figs. \ref{fig:low_eta1}(b) and \ref{fig:high_eta1}(b) show the evolution of the
ratio of mean energy densities of the total and uniform magnetic fields in the
disk $U_{\rm Bt}/U_{\rm Bu}$ for the low and high resolution respectively.
During the period of the first 500 Myr, the ratio is 1.0 in both cases, which
means that the turbulent magnetic component is negligible. Then the ratio starts
to rise very rapidly to the maximum value of about 1.15 and later on it
decreases. We can notice that values of $U_{\rm Bt}/U_{\rm Bu}$ are strongly
dependent on the uniform resistivity. The experiments with the low and high
number of the grid points show that both maximum values and the position of the
curve maxima are different for models with different values of the resistivity
coefficient $\eta_1$. The evolution of $U_{\rm Bt}/U_{\rm Bu}$ in models with
low resolution and with the lowest resistivity (p0, q0) have maximum values of
1.18 at the position of 1100 Myr. Increasing $\eta_1$ to 1.0 (r0) and 10.0 (s0)
lowers the maximum value to 1.13 but positions of the maxima are different for
these two cases: r0 has its maximum at about 1200 Myr, while s0 at 900~Myr. We
can see that there is no rule that increasing the resistivity causes a shift of
the maximum position to the right or the left side, both are possible. The last
experiment with the highest resistivity t0 ($\eta_1=100.0$) has a maximum of
about 1.025 more or less at the same time step as for the previous
experiment. The high resolution simulations result in a slightly different
evolution of the ratio $U_{\rm Bt}/U_{\rm Bu}$. Due to the lower value of the
numerical resistivity, the maximum for the model Q0 ($\eta_1=0.1$, see Table
\ref{tab:models}) is slightly higher than for the case P0 without physical
resistivity at all. The maxima for both curves are positioned at about 950 Myr,
so shifted to the left in comparison with the low resolution calculations. The
run R0 shows two maxima (about 900 and 1100 Myr) and then the evolution is
slightly different increasing its value until the end of the evolution. The
experiments with the higher resistivity coefficient (S0 and T0) are similar to
models with the low number of grid points (s0, t0, Fig. \ref{fig:low_eta1}(b).
This means that really high physical resistivity results in that the numerical
one is insignificant.

In case of uniform resistivity, plots of Fig. \ref{fig:low_eta1}(c) and Fig.
\ref{fig:high_eta1}(c) are presented to ensure of compatibility with the case of
the current-dependent resistivity discussed in the next Subsection.

The evolution of $j_{\rm max}$ (Fig. \ref{fig:low_eta1}(d),
\ref{fig:high_eta1}(d) shows similar character as the ratio $U_{\rm Bt}/U_{\rm
Bu}$. In the beginning the maximum of the current density is very small (about
0.03), but after 500 Myr it starts to grow rapidly to the maximum value of about
0.2 and 0.5 for the low and high resolution, respectively. This difference is
caused by the construction of the current density calculations, which depends on
a size of the grid step. In time less than 200 Myr, $j_{\rm max}$ reaches a
maximum, which is very dependent on the uniform resistivity. The evolution of
the maximum value of the current density in computations with low grid point
number gives a very similar value of the curve maxima (0.2) for all cases,
except the case t0 with $\eta_1=100.0$ showing only 0.03. The simulations with
high resistivity show a larger dispersion of maximum values, three cases (P0,
Q0, R0) have maximum value between 0.4 and 0.5, while the experiment S0
($\eta_1=10.0$) has 0.27 and T0 ($\eta_1=100.0$) has 0.03. In the saturation
phase, the value of $j_{\rm max}$ decreases. The inclination of all curves in
this phase is independent of the uniform resistivity for $\eta_1 < 10.0$ (runs
p0, q0, r0), but for the higher one some nonlinear effects may play a
significant role and the inclination is different. For example, the case s0 with
$\eta_1=10.0$ (see Fig. \ref{fig:low_eta1}(d)) contains three sub-phases in the
saturation state: from time 700 Myr, 900 Myr and from 1250 Myr, where the
inclination is different.

The deviation of magnetic field lines between opposite XZ boundaries is
presented in Figs. \ref{fig:low_eta1}(e), \ref{fig:low_eta1}(f) and Figs.
\ref{fig:high_eta1}(e), \ref{fig:high_eta1}(f) for the whole set of runs (see
Table 1) corresponding to the constant resistivity. At the end of the
exponential growth of perturbations the field lines disperse quickly, but later
on the resistivity limits further dispersion. We separated the standard
deviation of magnetic field lines into the X direction $\sigma_{\rm x}$ (Figs.
\ref{fig:low_eta1}(e), \ref{fig:high_eta1}(e)) and Z direction $\sigma_{\rm z}$
(Figs. \ref{fig:low_eta1}(f), \ref{fig:high_eta1}(f)). For both resolutions, the
magnetic field lines are dispersed more strongly in the Z direction than in
X direction, depending on the magnitude of the coefficient $\eta_1$. This can be
explained by the dominance of the vertical gravity force over the horizontally
directed Coriolis force. This relation between basic forces leads to vertical
current sheets as displayed in Fig. \ref{fig:slices}(c). The vertical current
sheets lead to the dominating vertical dispersion of reconnected magnetic field
lines. The maximum values of $\sigma_{\rm x}$ for experiments with zero or the
smallest resistivity (p0, q0, P0 and Q0) are about 80 pc for the high resolution
models and about 100 pc in the case of the low resolution runs. Values of
$\sigma_{\rm z}$ are much larger: 250 pc for the high resolution and about 150
pc for the low resolution runs. This behavior is consistent with the expectation
that lower resolution of simulations introduces larger numerical
resistivity. The maximum values of both dispersions decrease strongly with
increasing resistivity, resulting in 50 pc dispersion for the case t0 and T0
(resistivity 100.0, see Table \ref{tab:models}). Much smaller displacements
$\sigma_{\rm x}$ in the horizontal direction are related to the fact that the
primary force responsible for the instability is the vertical buoyancy force.

\subsection{Models with different values of $\eta_2$}
\label{sec:results_localized}

\begin{figure*}
  \centering
  \resizebox{0.7\hsize}{!}{\includegraphics{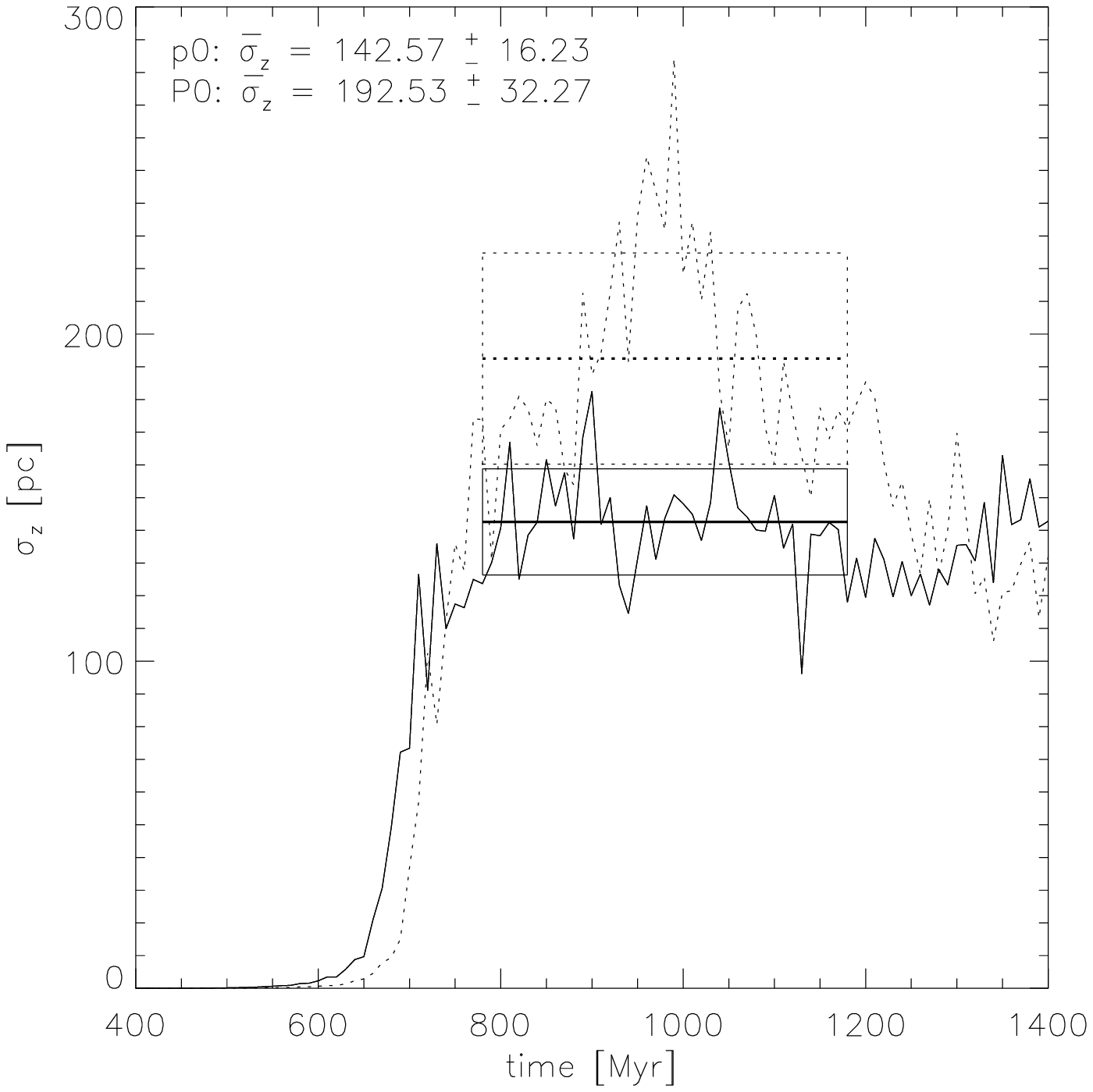}
                           \includegraphics{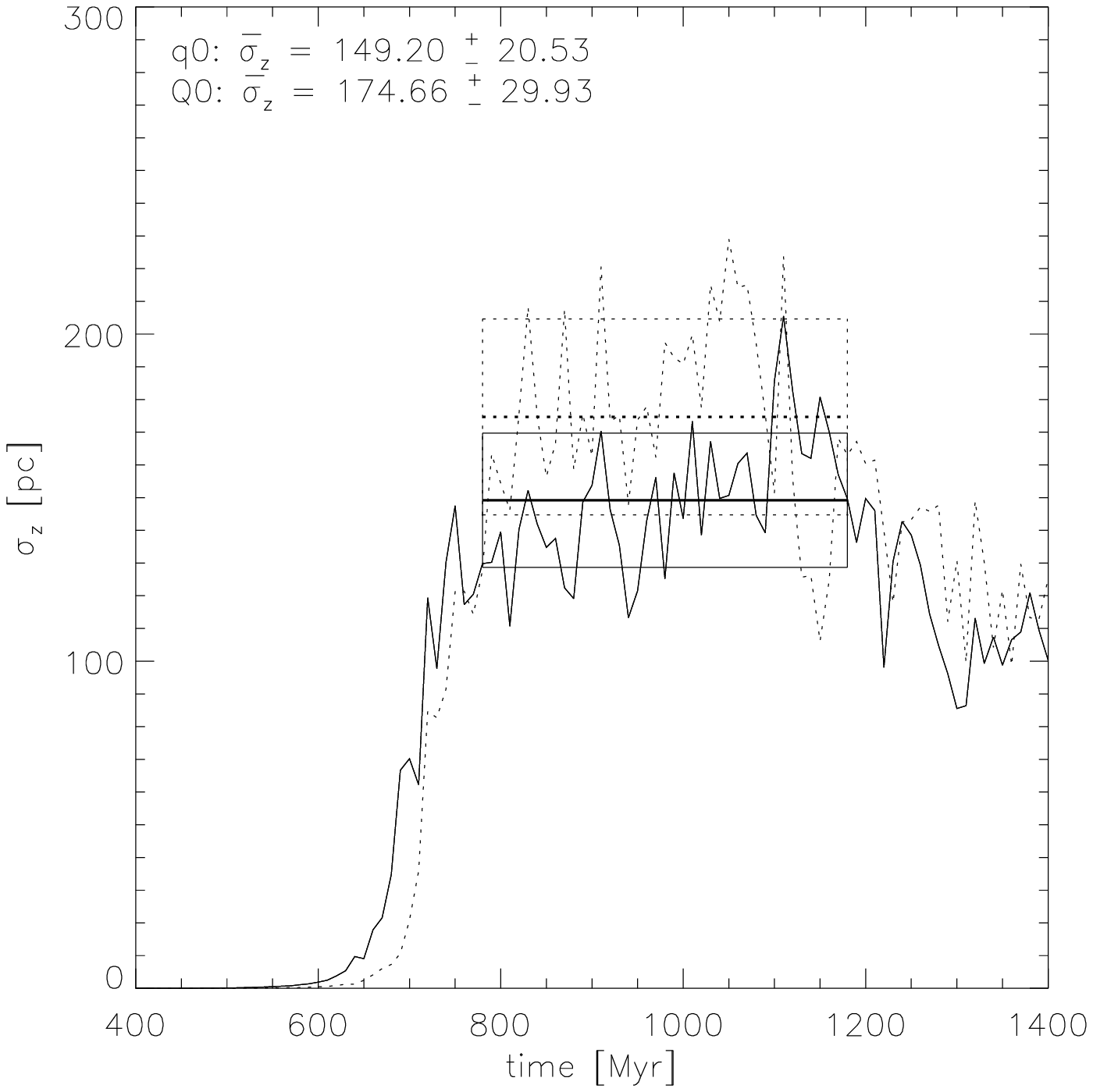}}
  \resizebox{0.7\hsize}{!}{\includegraphics{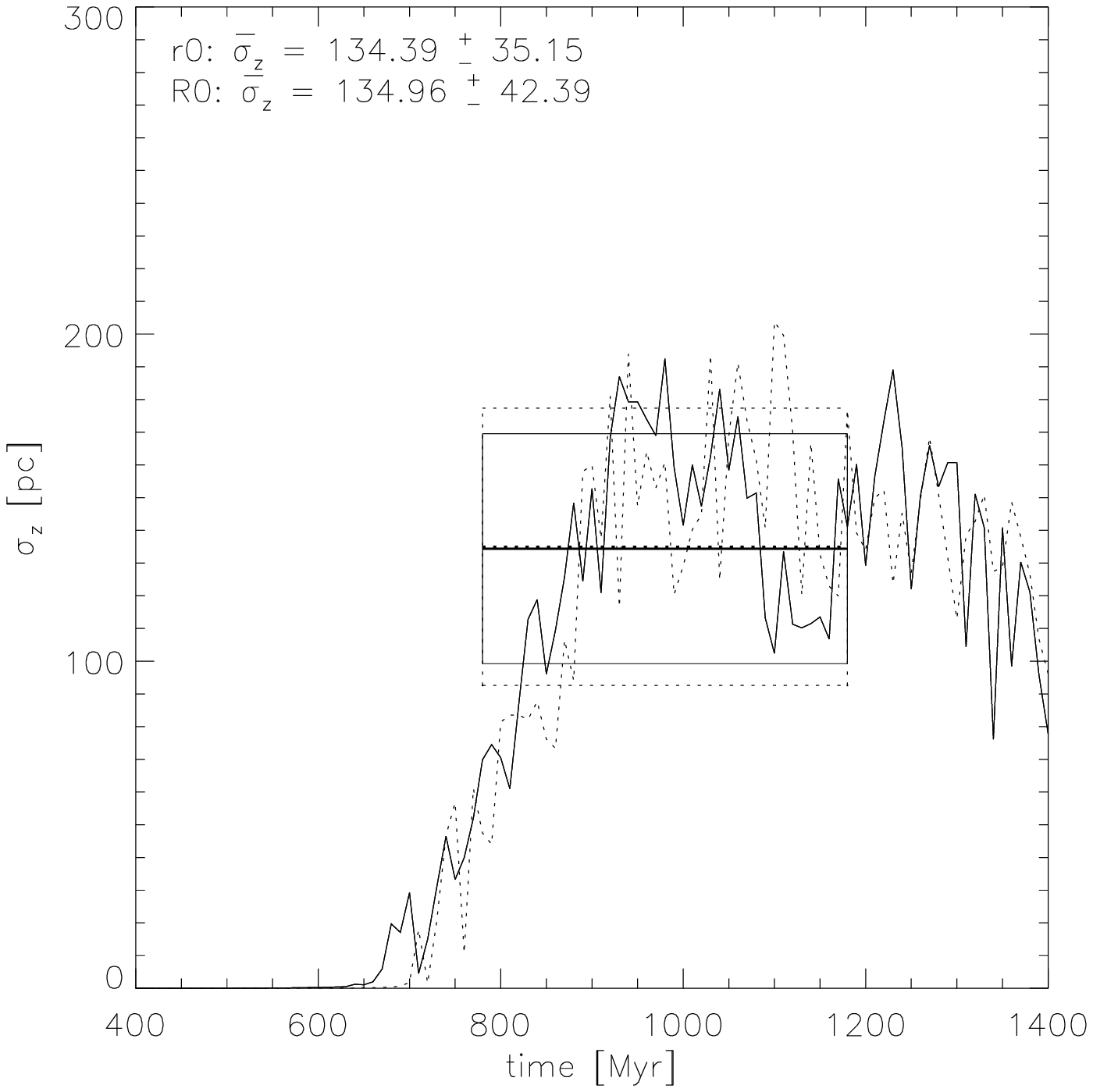}
                           \includegraphics{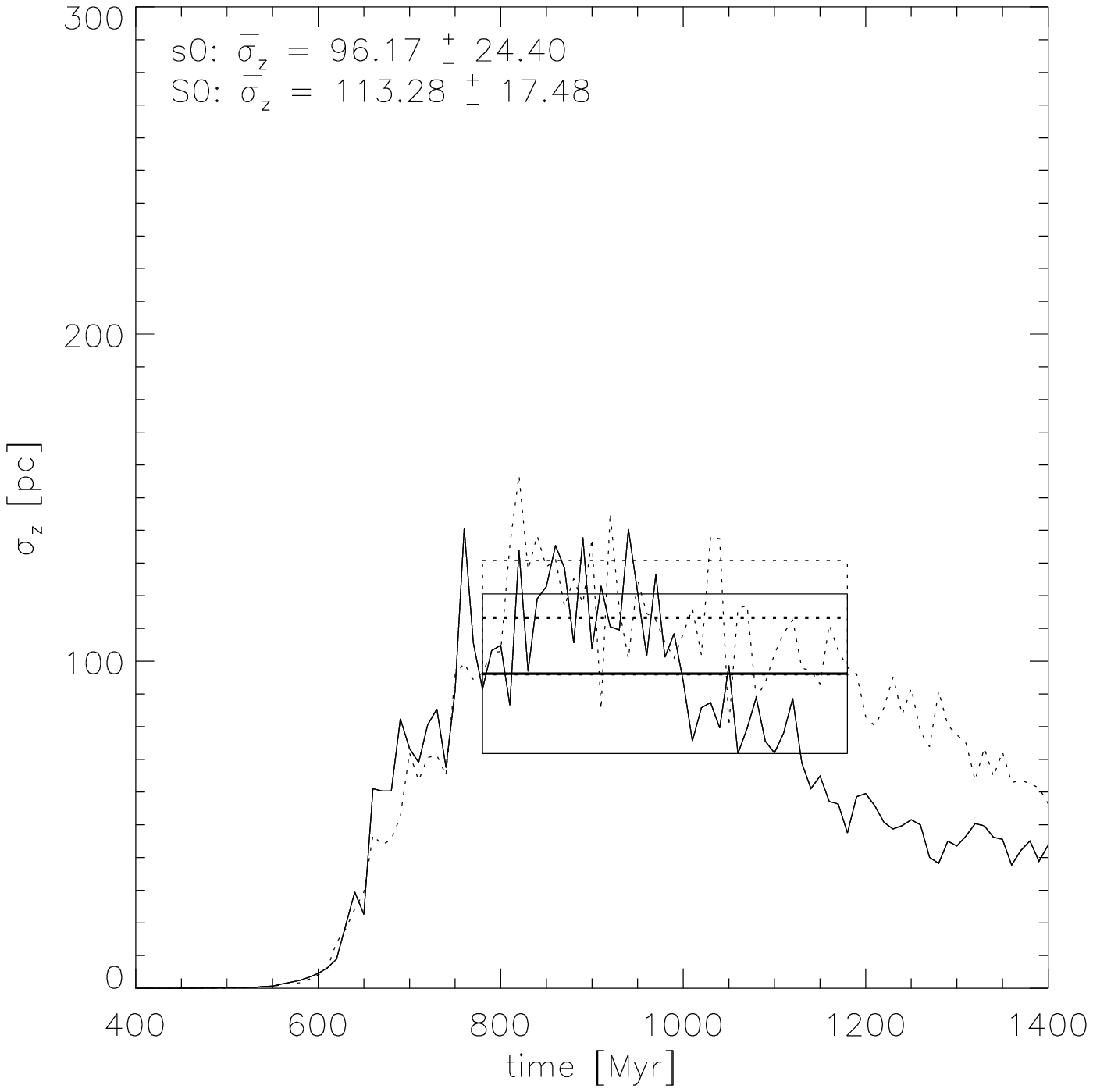}}
  \caption{Averaging of standard deviation in Z direction near maximum for
  models with different values of $\eta_1$. Thick lines indicate region of
  averaging and computed mean value of $\bar{\sigma_{\rm z}}$. Boxes around
  lines indicate errors. (solid - low resolution, dashed - high resolution)}
  \label{fig:plots_comp}
\end{figure*}

Figs. \ref{fig:low_eta2} and \ref{fig:high_eta2} present the time evolution of
our six statistical quantities in simulations with different values of $\eta_2$
(the low resolution cases: p0, p1, p2, p3, p4 and the high ones: P0, P1, P2, P3,
P4, see Table \ref{tab:models}). Generally, the influence of the localized
resistivity on the statistical quantities (a)-(f) is much weaker (except
$\eta_{\rm max}$) than the influence of the uniform resistivity.

The time evolution of $v_{\rm z \, max}$ proceeds similar to the case of the
uniform resistivity. In the present case, however, the evolution of the
exponential growth phase is identical for all runs. This is because the high
currents, localized in current sheets, starts to exceed the critical current
(may change for $j_{\rm crit}=0.1$) in the middle of the exponential growth
phase.  In the saturation phase, the maximum of $v_{\rm z}$ falls down less than
in the uniform resistivity case.

Comparing to the uniform resistivity (Figs. \ref{fig:low_eta1}(b) and
\ref{fig:high_eta1}(b)) the evolution of $U_{\rm Bt}/U_{\rm Bu}$ (Figs.
\ref{fig:low_eta2}(b) and \ref{fig:high_eta2}(b)) looks different especially for
the high resolution cases. This quantity exhibits a strong dependence on
$\eta_1$ but the effect of localized resistivity is only moderate. Now, the
ratio $U_{\rm Bt}/U_{\rm Bu}$ reaches comparable values for both grid
resolutions, while in the case of the uniform resistivity the resolution
influences the maximum values significantly.

For the high resolution cases $j_{\rm max}$ grows to large values, especially
for smaller values of $\eta_1$ (Figs.~\ref{fig:low_eta2}(d) and
\ref{fig:high_eta2}(d)). For example, for $\eta_1=0.0$ and $\eta_2=0.0$ $j_{\rm
max}$ reaches the value of 0.2 and 0.5 for the low and high resolution,
respectively. This difference we discussed in Subsection
\ref{sec:results_uniform}. For a large $\eta_2=100.0$ the evolution of $j_{\rm
max}$ rises only to the value of 0.2 i.e. exceeding the critical current much
less than in the case of small $\eta_2$.

The standard deviations of the magnetic field lines $\sigma_{\rm x}$ and
$\sigma_{\rm z}$ (Figs. \ref{fig:low_eta2}(e), \ref{fig:low_eta2}(f) and
\ref{fig:high_eta2}(e), \ref{fig:high_eta2}(f) do not exhibit an apparent
dependence on the resistivity at the small grid resolution, however at the high
resolution the behavior is opposite with respect to the uniform resolution. Up
to $t \sim 900$ Myr all three kinds of the resistivity produce the same
dispersion of magnetic field lines. Contrary to the other cases we note a
systematic growth of $\sigma_{\rm x}$ in the case of $\eta_2=100$ for the high
resolution. The dispersion of magnetic field lines in the X direction reaches
about 100 pc at $t=1500$ Myr, while $\sigma_{\rm x}$ is only 60 pc for
$\eta_2=0$. This property makes a significant difference in the localized
resistivity with respect to the uniform resistivity, because it is apparent that
the localized one does not tend to eliminate the initial dispersion of magnetic
field lines, which appeared in early phases of operation of the resistivity in
the period between $t=600$ and $t=900$ Myr.

\subsection{Determining of the numerical resistivity}
\label{sec:results_resistivity}

In our experiments we try to roughly estimate the value of the numerical
resistivity. The Parker instability does not grow infinitely but is suppressed
by the resistivity after exceeding some level, manifested by reaching the
saturation phase in the evolution of our six statistical quantities (see Section
\ref{sec:results}). The values of these quantities do not rise later on, but
after reaching some maximum point, oscillate around it, or decrease
exponentially. As an indicator of the total diffusivity in our models, we have
chosen the dispersion $\bar{\sigma_{\rm z}}$ of magnetic field lines defined in
Section 2.2. We average in time the values of this quantity near maximum,
i.e. the averaging is done for a period of about 400 Myr around maximum (see
Figure \ref{fig:plots_comp}). The computed values of $\bar{\sigma_{\rm z}}$ are
listed together with errors $\Delta \bar{\sigma_{\rm z}}$ in Table
\ref{tab:num_res}.

\begin{table}[t]
 \caption{\small Mean values of maximum of standard deviations in Z direction
                 for different models averaged from 800 Myr to 1200 Myr.}
 \begin{center}
  {\small
  \begin{tabular}{|c|c|c|c|}
   \hline Model & $\eta_1$ & $\bar{\sigma_{\rm z}}$ & $\Delta \bar{\sigma_{\rm
z}}$ \\ \hline \hline
   p0 &   0.0 & 142.6 & 16.2 \\
   q0 &   0.1 & 149.2 & 20.6 \\
   r0 &   1.0 & 150.0 & 25.3 \\
   s0 &  10.0 & 104.5 & 20.4 \\ \hline
   P0 &   0.0 & 192.5 & 32.3 \\
   Q0 &   0.1 & 174.7 & 30.0 \\
   R0 &   1.0 & 151.4 & 25.6 \\
   S0 &  10.0 & 112.6 & 21.0 \\ \hline
  \end{tabular}
  }
 \end{center}
 \label{tab:num_res}
\end{table}

For the uniform resistivity models p0, q0, and r0 (see Table \ref{tab:num_res},
low resolution), $\bar{\sigma_{\rm z}}$ has the same value of about $150.0 \pm
\Delta \bar{\sigma_{\rm z}}$ pc. Independently of the magnitude of explicit
resistivity, the time evolution of $\sigma_{\rm z}$ is similar for the runs p0,
q0, and r0. This fact indicates that numerical resistivity is large compared to
$\eta_1$. For $\eta_1 = 10.0$ (model s0) the situation is different. The
explicit resistivity is now dominant.

Comparing all the above with the high resolution models (P0, Q0, R0, and S0) we
conclude that the numerical diffusion in high resolution is much smaller,
because $\bar{\sigma_{\rm z}}$ is very large for the model with no explicit
resistivity and its value decreases with growing $\eta_1$.

\begin{figure}[t]
  \centering
  \resizebox{0.9\hsize}{!}{\includegraphics{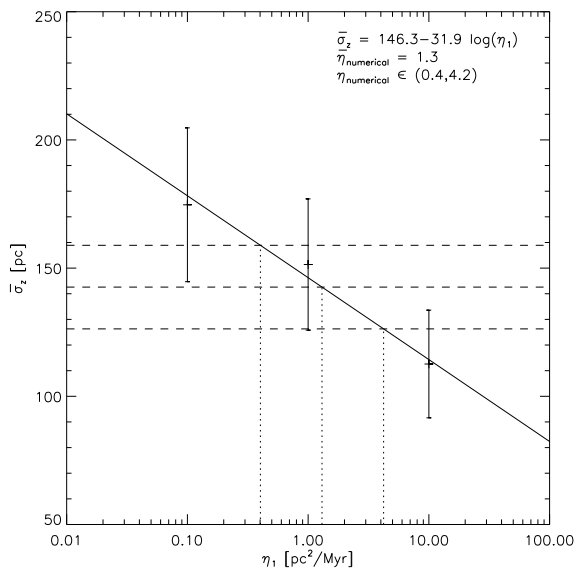}}
  \caption{Interpolation of $\eta_{\rm numerical}$.}
  \label{fig:estimation}
\end{figure}

Although we have very few points, we can roughly determine the numerical
resistivity in low resolution through plotting a log-lin graph of dependency
$\bar{\sigma_{\rm z}} = A + B \log( \eta_1 )$ for high resolution points (see
Fig. \ref{fig:estimation}). We obtain values $A=146.3$ and $B=-31.9$. Now we
take the value of $\bar{\sigma_{\rm z}}=142.6$ for model p0 (the low resolution,
no resistivity) and we project it onto our the curve. We can read the value of
$\eta_1 = 1.3$ which corresponds to $\bar{\sigma_{\rm z}}=142.6$. This means
that to get evolution of the Parker instability in the high resolution domain
similar to model p0 in the low resolution, we have to set $\eta_1 = 1.3$,
because this value corresponds to numerical resistivity in low resolution. It is
in agreement with our rough estimate above.

We can go further and include the error $\Delta \bar{\sigma_{\rm z}} = 16.2$ in
our estimations. By going through the same procedure as described above, but now
for values $\bar{\sigma_{\rm z}} + \Delta \bar{\sigma_{\rm z}} = 158.8$ and
$\bar{\sigma_{\rm z}} - \Delta \bar{\sigma_{\rm z}} = 126.4$, we estimated a
possible range for the value of the numerical resistivity, which is $\eta_{\rm 1
\, numerical} \in (0.4,4.2)$ (see Fig. \ref{fig:estimation}).

The procedure described above does not exclude the numerical resistivity in the
high number of grid points experiments. After performing a set of simulations in
higher resolutions we should be able to plot a dependency of the numerical
resistivity on the resolution. Then it will be possible to determine the value
of $\eta_{\rm 1 \, numerical}$.

\section{Discussion}
\label{sec:discussion}

The central issue of the present project is to identify the effects of three
kinds of resistivity (uniform, current dependent and numerical) on the
topological evolution of magnetic field lines after the onset of the Parker
instability. We found that the numerical resistivity present in the
currently-used code Zeus3D produces dispersion of magnetic field lines, which is
present in the action of a weak uniform resistivity.

This finding allowed us to propose a new method of determination of the
numerical diffusion in the numerical simulations of the Parker instability. The
method described in Subsect. \ref{sec:method_measure} is based on a statistical
analysis of the dispersion of magnetic field lines in a non-ideal medium.

We measure the dispersion of a large set of magnetic field lines and relate this
dispersion to the type and the magnitude of the resistivity. This kind of
analysis is applicable to MHD simulations in a local approximation involving
periodic boundary conditions at least on two planes which are intersected
simultaneously by a family of magnetic field lines.

We determine the numerical resistivity for low resolution simulations comparing
the mean dispersion of magnetic field lines in the case of the vanishing
physical resistivity to the case of high resolution and several values of the
uniform resistivity.

The evolution of our statistical quantities shows that for the largest values of
the resistivity and higher resolution of simulations the numerical resistivity
is negligible. In some cases the extremely large uniform resistivity impedes the
Parker instability.

Our considerations imply that the Parker instability contributes to a
randomization of galactic magnetic fields. The main sources of perturbations in
real galaxies are supernovae supplying locally kinetic, thermal and cosmic ray
energy. These perturbations are the most plausible mechanism of the excitation
of the Parker instability and the randomization of the magnetic field in spiral
arms. The resistivity is then expected to contribute to the annihilation of the
random magnetic field component in the inter-arm regions. In galaxies like
NGC6946 and M51 the magnetic field appears to be much more uniform in the
inter-arm regions then in arms (see model of Rohde et al. \cite{rohde}).

The modeling of the Parker instability in the present simulations is far from
being realistic. In order to make our model feasible we applied simplifying
assumptions of a plane-parallel initial magnetic field, uniform vertical
gravity, isothermal evolution of the thermal gas, a lack of cosmic rays, a lack
of selfgravity and differential rotation in the disk. All these circumstances
imply that numbers and timescales of different phenomena are not realistic
too. Among different effects we can expect that the presence of cosmic rays
significantly shortens all the relevant timescales of observed phenomena,
however we believe that future incorporation of all the complexity of realistic
galactic disks will not change the qualitative picture of the evolution of a
magnetic field structure.

\section{Summary and conclusions}
\label{sec:summary}

In the present paper we considered the general problem of the influence of the
physical and numerical resistivity on the Parker instability. We focused on the
issue of the dispersion of magnetic field lines (which according to Subsect. 2.2
is understood in this paper in the topological sense), resulting from three
types i.e. uniform, current-dependent and numerical resistivities.

We performed several experiments applying two different grid resolutions and
different types and magnitudes of resistivity. We propose a new statistical
method of quantitative estimation of the influence of the explicit and the
numerical resistivity on the dispersion of magnetic lines and the amount of a
random component of the magnetic field in the considered MHD system.

Our conclusions are:
\begin{enumerate}

\item Except for the most extreme value of the uniform resistivity $\eta_1=100$,
the resistivity does not influence the growth rate of the Parker
instability. For extremal values of the resistivity, the linear growth rate of
Parker instability is significantly diminished.

\item The time evolution of the dispersion of magnetic field lines and the ratio
of the total to uniform magnetic field is strongly dependent on the type and the
magnitude of resistivity.

\item The presence of all kinds of resistivity leads to a rapid dispersion of
magnetic field lines at the end of the exponential growth phase of evolution of
the Parker instability. After the initial sudden dispersion of magnetic lines
the subsequent evolution depends on the type of resistivity.

- uniform resistivity tends to convert the disturbed magnetic field structure to
the uniform state (uniformization of the magnetic field). This effect is clearly
seen in the plots of the ratio of total to uniform magnetic energies and in the
plots of the dispersion of magnetic field lines.

- the localized resistivity diminishes the ratio of the total to uniform
magnetic energies less efficiently than the uniform resistivity. On the other
hand, for high resolution runs we observe the growth of the horizontal
dispersion of magnetic field lines in the course of time, while the vertical
dispersion tends to decrease.

- the effect of the numerical resistivity is dependent on the grid resolution
and is qualitatively similar to the effect of uniform resistivity. We determine
the magnitude of the uniform resistivity equivalent to the numerical resistivity
in simulations with lower resolution.

\item The method can be applied to compare the numerical magnetic diffusivity of
different codes and studies of different physical processes e.g. turbulence.

\end{enumerate}

\begin{acknowledgements}
We thank Polish Committee for Scientific Research (KBN) for the support through
the grants PB 4264/P03/99/17 and PB 249/P03/2001/21.
\end{acknowledgements}

\end{document}